# Super-hard and superconducting boron clathrates in the prediction of U-B compounds

Juefei Wu[1], Dexi Shao[2], Junjie Wang[3], Yu Han[3], Bangshuai Zhu[1], Cuiying Pei[1], Qi Wang[1,4], Jian Sun[3*] and Yanpeng Qi[1,4,5*]

1. State Key Laboratory of Quantum Functional Materials, School of Physical Science and Technology, ShanghaiTech University, Shanghai 201210, China
2. School of Physics, Hangzhou Normal University, Hangzhou 311121, China
3. National Laboratory of Solid State Microstructures, School of Physics and Collaborative Innovation Center of Advanced Microstructures, Nanjing University, Nanjing 210093, China
4. ShanghaiTech Laboratory for Topological Physics, ShanghaiTech University, Shanghai 201210, China
5. Shanghai Key Laboratory of High-resolution Electron Microscopy, ShanghaiTech University, Shanghai 201210, China

* Correspondence should be addressed to Y.P.Q. (qiyp@shanghaitech.edu.cn) or J.S. (jiansun@nju.edu.cn).

## Abstract

The binary metal borides provide a promising platform for searching unique materials with superconductivity and super-hardness under high pressure, owing to the distinctive bonding characters of boron. In this work, combined the first-principles calculations and crystal structure predictions, we predicted 4 exotic stoichiometries and 8 unique U-B compounds under high pressure. The predicted compounds have layered or caged structure units and 4 of them host high hardness under ambient pressure. By removal of the U atoms, we predicted three meta-stable boron clathrates at ambient pressure. Remarkably, the Vickers hardness of the predicted $C2/m$-$B_6$ is estimated to be 49-53 GPa, and the $C2/m$-$B_{12}$ is superconducting with the $T_c$ value of 16.12 K. Our calculations enrich the phase diagram of binary metal borides and boron allotropes, providing insights for the future theoretical and experimental studies on unique materials.

# I. Introduction

Binary metal borides have a significant variety of morphologies including 3D polyhedron, 2D nets and 1D chains, stemming from the valence electrons of boron are insufficient to fill up the bonding states and tend to form covalent bonds with multi-centers [1-3]. The combination of the boron based structural units and different metal elements are fundamental to the fascinating properties of binary metal borides, including the superconductivity and mechanical properties [4], thus has gained tremendous research attention. For instance, the superconducting $MgB_2$ has a high transition temperature $T_c$ of 39 K at ambient pressure, which ignites the investigations of exotic superconductors in borides [5, 6]. Recent study reports that $α$-$MoB_2$ has the highest $T_c$ in the transition metal borides with the value of 32.4 K at 109.7 GPa [7, 8]. Subsequently, the $MgB_2$-like $WB_2$ becomes superconducting under high pressure with a $T_c$ of ~17 K at 90 GPa [9, 10]. The Mo-$d$ and W-$d$ orbitals play crucial roles in the superconductivity of corresponding borides. In the theoretical aspect, $CaB_2$ is iso-structural to $MgB_2$ and hosts stronger electron-phonon coupling (EPC) with $T_c$ of 48 K [11]. The predicted $YB_6$ composes of $B_{24}$ units and is superconducting with a $T_c$ of 12.78 K at ambient pressure [12]. Ref. 13 predicts $LaB_8$ consisting of $B_{26}$ cages, which is thermodynamically stable above 70 GPa and the $T_c$ increases to ~20 K at ambient pressure.

In addition to superconductivity, binary metal borides are promising for exploring super-hard materials (hardness ≥ 40 GPa) [14, 15] due to the strong covalent B-B bonds. The $ReB_2$ comprises boron layers and the hardness reaches 48 GPa in the experiment [16]. $WB_4$ contains hexagonal B layers and B dimers, proposed with the hardness of 43.3 GPa [17]. The stackings of $FeB_{12}$ polyhedra columns construct $FeB_4$, which leads to hardness of 62 GPa [18]. $ZrB_{12}$ is made up of the $B_{24}$ units with the hardness of 40 GPa [19]. Besides, theoretical works predict excellent hard and super-hard binary metal borides with diverse boron structure motifs. The predicted $CrB_4$ encompasses interconnected quadrilateral boron units with the hardness of 48 GPa [20]. $WB_5$ holds boron networks with the estimated hardness of 45 GPa [21]. The Ref. 22 predicts that $ScB_3$ contains $ScB_{13}$ polyhedron and $ScB_6$ involves boron trigons, and their calculated hardness values are of 38.3 GPa and 39.8 GPa, respectively.

Meanwhile, the high hardness and superconductivity could coexist in binary metal borides as well. The predicted $α$-$BeB_6$ is a super-hard material (46 GPa) with $T_c$ of 9 K, and the high pressure phase $β$-$BeB_6$ has hardness of 31 GPa and $T_c$ of 21 K [23]. Ref. 24 predicts a class of hard superconductors constituted of $B_{23}$ cages, such as

KB$_7$ (hardness 22.5 GPa and $T_c$ 26.2 K) and SrB$_7$ (hardness 25.1 GPa and $T_c$ 12.67 K). Ref. 25 proposes the hardness values of clathrate structures LiB$_4$ and NaB$_4$ are about 39 GPa, and their $T_c$ values are of 6 K and 8 K, respectively. Recent study on clathrate-like CeB$_n$ compounds report the hardness of 20-39 GPa, and $T_c$ values are of 29 K and 27 K for CeB$_2$ for CeB$_8$, respectively [26].

Moreover, binary metal borides can act as precursors to achieve boron allotropes with high hardness and superconductivity. The clathrate NaB$_4$ and Na$_2$B$_{17}$ are the precursors of B$_4$ ($T_c$ 19.8 K) and B$_{17}$ ($T_c$ 15.4 K), respectively [27]. Ref. 28 enhances the $T_c$ record in elemental superconductors and predicts caged $t$-B with the $T_c$ value of 43 K at ambient pressure, which is much higher than its precursor $t$-MnB$_{12}$ ($T_c \sim 7$ K) formed by B$_{16}$ cages. The predicted $o$-B$_{16}$ exhibits $T_c$ of 14.2 K and hardness of 19.4 GPa, which originates from the removal of Sr atoms in the SrB$_8$ featuring anticlinal pentapyramids units [29].

Therefore, the above results suggest that both the unique boron morphologies and appropriate metal elements are key factors for pursuing either super-hard materials or superconductors in binary metal borides, which may also benefit for searching exotic boron allotropes with interesting properties. Inspiring by the researches of superconducting hydrides under high pressure, the element uranium could help to reduce the stable pressure and play crucial role in the electronic structures and EPC of hydrides [30-34]. Besides, the stoichiometries along with the layered or caged structure units of the known U-B compounds UB$_2$, UB$_4$ and UB$_{12}$ [35, 36] illustrate the potential to predict unique boron-rich structures under high pressure. In addition, UB$_2$ is also a potential candidate as safer fuel for Generation IV nuclear reactors [35], while the studies of U-B compounds under high pressure are relatively limited. Thus, it is intriguing to study the U-B compounds under high pressure. In this work, we explore the U-B compounds by combining the first-principles calculations and the machine learning graph theory accelerated crystal structure search method within 200 GPa. We predict 4 exotic stoichiometries and 8 unique U-B compounds, and 4 of them have high hardness under ambient pressure. By removing the uranium atoms, we predict 3 boron clathrates with meta-stability at ambient pressure. Among them, the Vickers hardness of the predicted $C2/m$-B$_6$ is estimated to be 49-53 GPa, and $C2/m$-B$_{12}$ is superconducting with the $T_c$ value of 16.12 K.

## II. Methods

Firstly, we conducted first-principles calculations to evaluate the influence of the on-site Coulomb repulsion effects on the localized 5$f$ electrons of U atoms in U-B compounds based on the experimental data under high pressure [35]. We used the Vienna *Ab-initio* Simulation Package (VASP) based on the density functional theory [37, 38], and the detailed settings are listed in Supplementary file [39]. The test results are shown in Table. S1 [39], the Coulomb repulsion has negligible impact on the lattice parameters, and the results match well with the experiment data without Coulomb repulsion under high pressure [39]. This is analogous to the small effect of Coulomb repulsion in U-H compounds under high pressure [31, 32].

Then we carried out the crystal structure searches by combining VASP with the machine learning and graph theory assisted universal structure searcher (MAGUS) [43, 44], conducting variable composition structure searches to construct the convex hulls. Fixed composition structure searches were further performed for each stable stoichiometry to validate the results. The exchange-correlation functional was the generalized gradient approximation (GGA) of Perdew, Burkey, and Ernzerhof [45], and valence electrons of $5f^36s^26p^66d^27s^2$ in U and $2s^22p^1$ in B are treated by the projector-augmented wave (PAW) approach [46] for geometry optimization and electronic structure calculations. To study the chemical bonding properties, we performed the crystal orbital Hamilton population (COHP) analysis using the LOBSTER 4.1 package with the pbeVaspFit2015 basis set [47]. The phonon spectra were calculated by the PHONOPY program package [48]. To further access the thermal stability, we conducted *ab initio* molecular dynamics (AIMD) simulations under *NVT* ensemble (*N*, *V*, and *T* are the number of particles, volume, and temperature, respectively) through the Nosé-Hoover approach [49], which is rechecked by *NPT* ensemble (*P* is pressure) as well. The elastic constants were calculated using the strain-stress method [50], and we employed the Voigt-Reuss-Hill approximation to derive the bulk modulus $B$ (GPa), shear modulus $G$ (GPa) and Young's modulus $E$ (GPa) [51]. The Vickers hardness ($H_v$) is estimated by Chen's formula ($H_v^{Chen}$) [52], Tian's formula ($H_v^{Tian}$) [53] and Mazhnik's formula ($H_v^{Mazhnik}$) [54]. The electron-phonon coupling (EPC) constants were calculated by the QUANTUM-ESPRESSO (QE) package [55]. More detailed parameters for the above calculations can be seen in the Supplementary material [39].

## III. Results and discussion

## A. Structures and stability of U-B compounds

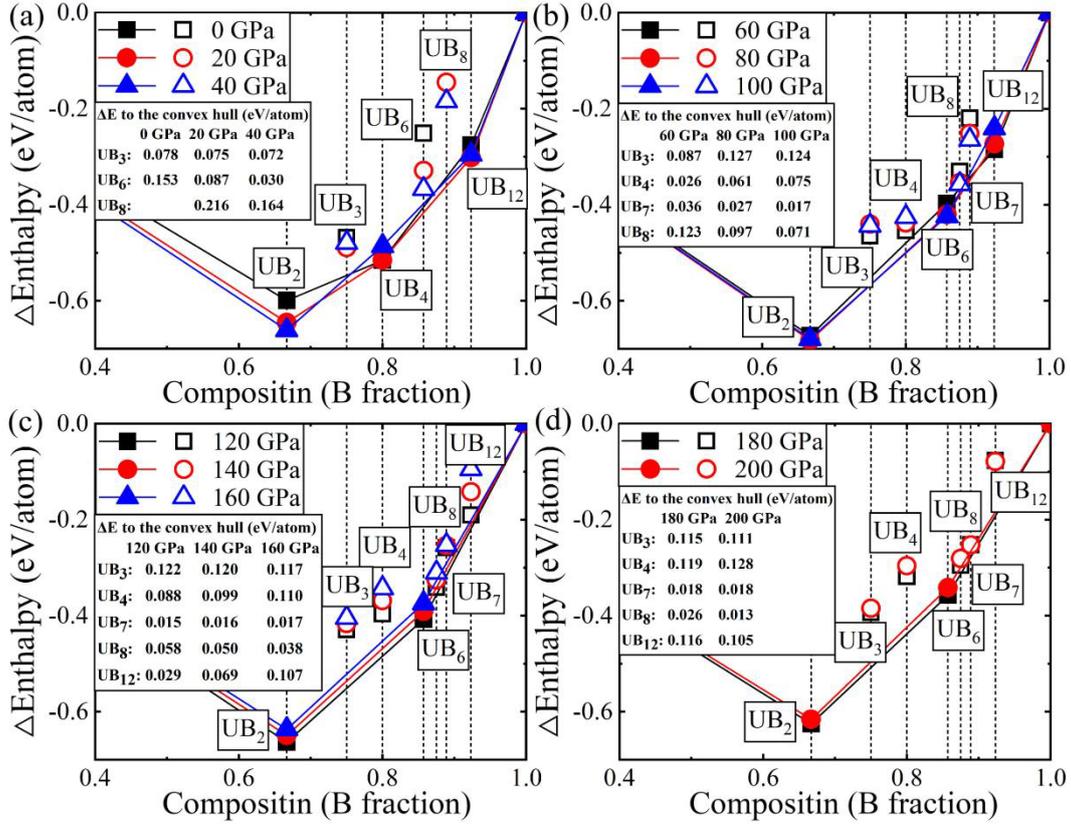

FIG. 1. The convex hull diagrams for U-B compounds under high pressure. The solid points indicate that the stoichiometries are on the convex hull, while the empty points are above the convex hull.

In this work, we explore the stable phases of U-B compounds under high pressure and conducted structure searches at 50, 100, 150 and 200 GPa, respectively. We recalculate the enthalpies of the predicted structures to construct the convex hull with the simple substances U and B [56-58] as the references under high pressure. As shown in Fig. 1, we predicted 4 unique stoichiometries $UB_3$, $UB_6$, $UB_7$ and $UB_8$. Since all the stoichiometries are on the boron-rich side, the convex hull starts from the B/(U+B)=0.5 site. In Fig. 1 (a), the known stoichiometries $UB_2$, $UB_4$ and $UB_{12}$ lie on the convex hull at 0 GPa, the consistency with the experimental results [35] suggest the feasibility of our calculation methods. $UB_2$ remains on the convex hull up to 200 GPa, while $UB_4$ falls off the convex hull after 60 GPa and the enthalpy difference relative to the convex hull increases with the pressure. Simultaneously, the predicted stoichiometry $UB_6$ emerges after 60 GPa and persists within the calculated pressure range. With further compression, the known stoichiometry $UB_{12}$ has analogous trend with $UB_4$, which leaves the convex hull since 120 GPa. Besides, the predicted stoichiometries $UB_3$, $UB_7$ and $UB_8$ approach the convex hull with pressure increasing, in particular UB$_8$ after 120 GPa.

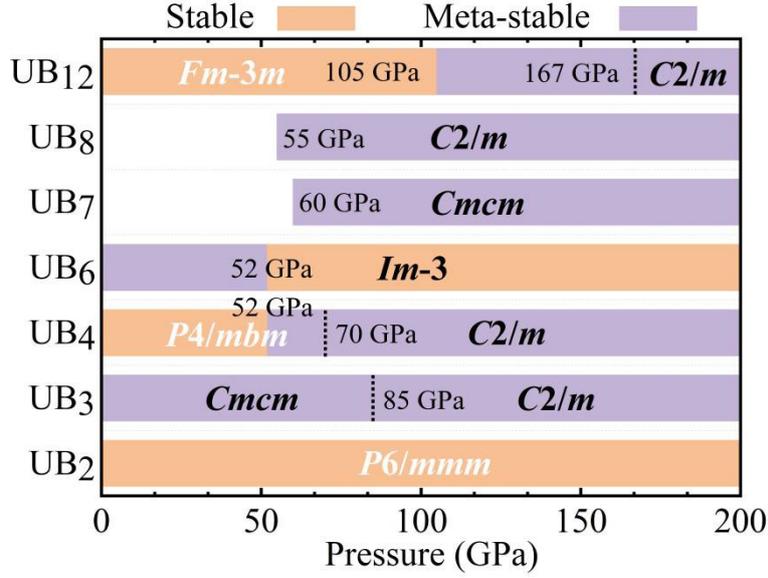

FIG. 2. The pressure-composition phase diagram of the U-B compounds. The predicted structures are in black bold font, while the reported structures are in white font. The orange region is the stable state and the violet region is the meta-stable state. The dashed lines suggest the phase transition pressure in the stoichiometry.

Then we calculated the detailed relative enthalpies [Fig. S1] [39] and the phonon spectra [Fig. S2-S3] [39] for the phase diagram of U-B compounds under pressure [Fig. 2]. To distinguish the thermodynamic meta-stability, a common threshold is of 50 meV/atom in the field of structures predictions [59], while the threshold of 200 meV/atom is employed for predictions in regard to the synthesized metal compounds [60]. Besides, the relative formation energy could reach 300 meV/atom in HfB [61], and the meta-stable Al$_{1.28}$B phase has positive formation energy around 208 meV/atom [62]. Hence we take the relatively intermediate threshold of 130 meV/atom for the meta-stability border anchoring the Mn-B compounds under high pressure [28]. The structure searches predicted 8 structures including UB$_3$-*Cmcm*, UB$_3$-*C2/m*, UB$_4$-*C2/m*, UB$_6$-*Im-3*, UB$_6$-*Cmcm*, UB$_7$-*Cmcm*, UB$_8$-*C2/m* and UB$_{12}$-*C2/m*, all of which have potential for synthesizing under high pressure. The calculated lattice parameters of the predicted structures are listed in Table. S2 [39]. The phonon spectra of the predicted structures have no imaginary frequencies [Fig. S2-S3] [39], indicating their dynamical stability within the corresponding pressure ranges. In addition, we calculated the zero point energy (ZPE) and estimated the temperature effect on the convex hull diagrams [Fig. S4] [39]. In Fig. S4 (a) and (d) [39], the enthalpy difference relative to the convex hull have slight changes, while the nodes of the convex hull and the thermodynamic stability keep consistent. Besides, the temperature effect has less effect on the convex hull diagrams [Fig. S4 (b) and (d)] [39]. Hence we believe that the effect of ZPE could be negligible. According to our calculations, both $UB_3$-*Cmcm* and $UB_3$-*C2/m* are thermodynamically meta-stable under high pressure [Fig. S1(a)] [39] and they are dynamically stable at ambient pressure [Fig. S2(a)-(d)] [39]. Because of the emergence of $UB_6$, the known structure $UB_4$-*P4/mbm* becomes meta-stable after 52 GPa and we predict a structural transition to $UB_4$-*C2/m* about 70 GPa [Fig. S1(b)] [39]. As plotted in the inset of Fig. S1 (c) [39], the predicted $UB_6$-*Im*-3 is accompanied with a meta-stable structure $UB_6$-*Cmcm*. Meanwhile, $UB_6$ also breaks the thermodynamic stability of $UB_7$ and $UB_8$, keeping them from the convex hull under high pressure [Fig. S1 (d) and (e)] [39]. As for $UB_{12}$ [Fig. S1(f)] [39], the known structure $UB_{12}$-*Fm*-3*m* enters the meta-stable state since 105 GPa and it transforms to the predicted $UB_{12}$-*C2/m* around 167 GPa. $UB_{12}$-*C2/m* is not dynamically stable at the ambient pressure, but becomes stable until the critical pressure of 40 GPa [Fig. S3(g)] [39].

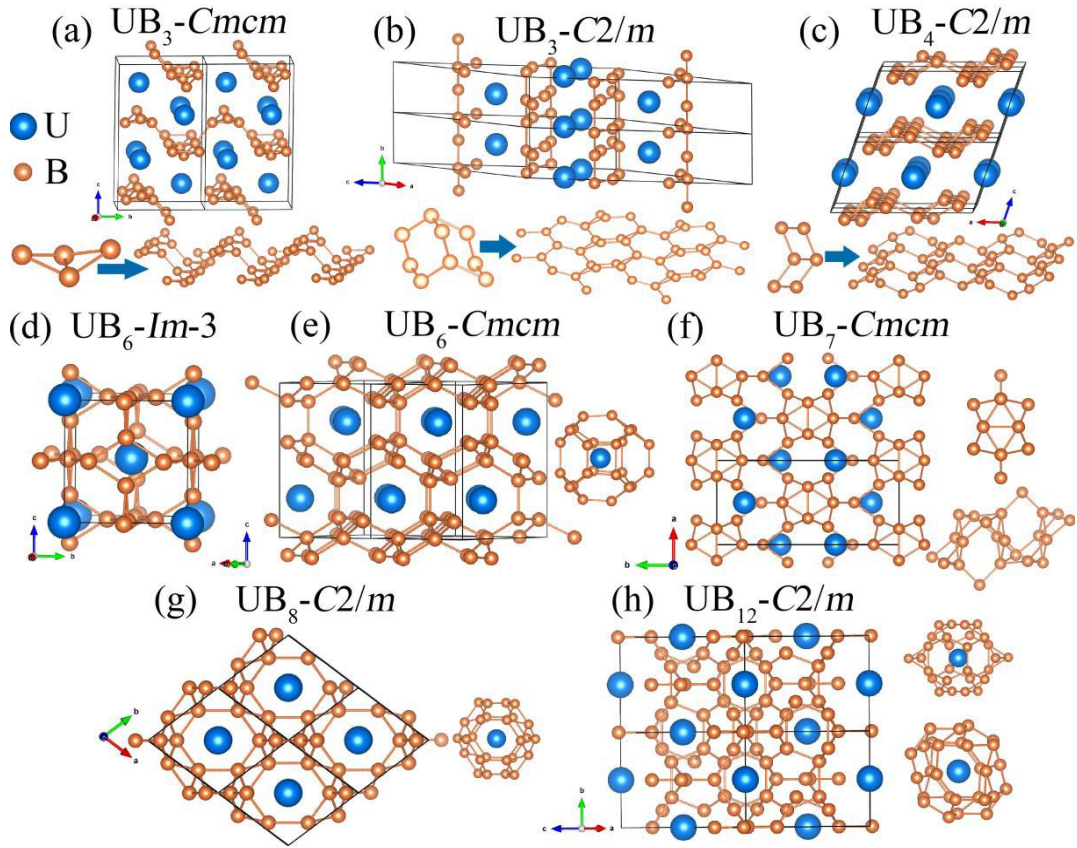

FIG. 3. The crystal structures of the predicted U-B compounds. The insets are the structure units.

As shown in Fig. 3 (a)-(c), the crystal structures of the predicted $UB_3$-*Cmcm*, $UB_3$-*C2/m* and $UB_4$-*C2/m* are constructed by the stacking of boron networks with the uranium atoms embedded between the networks. The basic unit in the predicted $UB_3$-*Cmcm* is the two-fold boron triangles [inset of Fig. 3(a)]. The units form the boron chains and the connection between the chains result in the waved boron layer in UB$_3$-*Cmcm*. The boron nets in the predicted UB$_3$-*C2/m* are composed of a two-fold six-member rings [inset of Fig. 3(b)], while the nets in the predicted UB$_4$-*C2/m* is composed of the four-member rings [inset of Fig. 3(c)]. We can observe flat hexagons between the structure units in both UB$_3$-*C2/m* and UB$_4$-*C2/m*. In Fig. 3 (d), the four-member and six-member rings of boron atoms construct a B$_{24}$ cage in the cubic UB$_6$-*Im*-3, resembling the hydrogen cages in the hydrides under high pressure [34]. Since the basic unit of UB$_6$-*Cmcm* [Fig. 3(e)] is the distorted B$_{24}$ cages of UB$_6$-*Im*-3, we can regard UB$_6$-*Cmcm* as the sliding of the UB$_6$-*Im*-3. In terms of the predicted UB$_7$-*Cmcm* [Fig. 3(f)], the boron atoms arrange in columns, with the uranium atoms embedding between boron columns. Both UB$_8$-*C2/m* and UB$_{12}$-*C2/m* exhibit clathrate features, in which the boron cages form channels and encircle the guest uranium atoms. The boron cages in UB$_8$-*C2/m* and UB$_{12}$-*C2/m* contain relatively distorted rings. The B$_{26}$ cage in the UB$_8$-*C2/m* [Fig. 3(g)] consists of four-member, twisted five-member and six-member rings, while the B$_{28}$ cage in the UB$_{12}$-*C2/m* [Fig. 3(g)] is composed of five-member and twisted six-member rings.

## B. Electronic structures and mechanical properties

We calculated the band structures and projected density of states (PDOS) of the predicted U-B compounds at their corresponding lowest dynamically stable pressure, as illustrated in Fig. S5 and S6 [39]. The valence bands and conduction bands overlap with each other around the Fermi energy in all the predicted U-B compounds, indicating typical metal features. Despite the contribution of B atoms to the PDOS around the Fermi energy enhances with the B content, such as in UB$_7$-*Cmcm* and UB$_{12}$-*C2/m*, the U-*f* electrons play dominant role in the PDOS around the Fermi energy in all the predicted U-B compounds. By calculating the electron localization function (ELF), we analysed the bonding types in B-B and U-B in the predicted U-B compounds. As shown in Fig. S7 [39], the bonds between B atoms indicates covalent interactions in the nets and cages. Nevertheless, we can find the connections between U atoms and B atoms, suggesting the U-B bonds are more covalent than ionic. It is similar with the interactions in binary U-H [30-32] and ternary U-Ca-H [34], in which the U-*f* electrons give rise to the metallicity in these hydrides. To further understand the bonding properties of the predicted U-B compounds, we performed the COHP and integrated COHP (ICOHP) calculations. According to the convention, the positive and negative values of -COHP represent bonding and anti-bonding characteristics and ICOHP could identify the bonding strength. As plotted in Fig. S8 [39], both bonds between B-B and U-B are overall localized in the positive region of -COHP below the Fermi energy, illustrating the bonding fingerprints of B-B and U-B in all the predicted U-B compounds. The ICOHP results are listed in Table. 1, the bond strength of B-B is not only stronger than U-B in U-B compounds, but also stronger than the B-B bonds proposed in NaB$_4$ [25], MnB$_{12}$ [28], SrB$_8$ [29] and is comparable to the boron allotrope $o$-B$_{16}$ [29]. Furthermore, the ICOHP absolute value of U-B bonds is larger than that of Na-B, Mn-B and Sr-B [25, 28, 29], which accords with the ELF feature in Fig. S7 [39]. The stronger bond strength implies the potential of high hardness in the predicted U-B compounds. Besides, The Mulliken charge analysis demonstrate that the U atoms at most transfer 1.50 e$^-$/atom to B atoms in UB$_6$-$Im$-3 and UB$_6$-$Cmcm$, and we find that the bond distances have almost negative correlation with the bond strength in U-B. Moreover, we also calculated the EPC properties of the predicted U-B compounds, only UB$_6$-$Cmcm$ and UB$_8$-$C2/m$ are possible superconductors with the EPC constants $\lambda$ 0.30 and 0.35, suggesting weak superconductivity with estimated $T_c$ values < 1 K.

TABLE. 1. The averaged ICOHP, averaged bond length and charge transfers of the predicted U-B compounds.

|  | Pressure (GPa) | B-B ICOHP (eV/pair) | U-B ICOHP (eV/pair) | B-B bond (Å) | U-B bond (Å) | B charge (e$^-$/atom) | U charge (e$^-$/atom) |
|---|---|---|---|---|---|---|---|
| UB$_3$-$Cmcm$ | 0 | -5.58 | -1.65 | 1.77 | 2.73 | -0.27 | 0.82 |
| UB$_3$-$C2/m$ | 0 | -5.40 | -1.95 | 1.78 | 2.70 | -0.33 | 1.01 |
| UB$_4$-$C2/m$ | 0 | -5.79 | -1.89 | 1.74 | 2.69 | -0.28 | 1.13 |
| UB$_6$-$Im$-3 | 0 | -5.53 | -1.55 | 1.76 | 2.74 | -0.25 | 1.50 |
| UB$_6$-$Cmcm$ | 0 | -5.79 | -1.52 | 1.75 | 2.74 | -0.25 | 1.50 |
| UB$_7$-$Cmcm$ | 60 | -5.53 | -2.47 | 1.70 | 2.52 | -0.14 | 0.97 |
| UB$_8$-$C2/m$ | 20 | -5.55 | -2.41 | 1.74 | 2.57 | -0.17 | 1.36 |
| UB$_{12}$-$C2/m$ | 40 | -5.62 | -1.65 | 1.70 | 2.70 | -0.12 | 1.40 |

Furthermore, we calculated the mechanical properties of the predicted U-B compounds including bulk modulus $B$ (GPa), shear modulus $G$ (GPa), Young's modulus $E$ (GPa) and Vickers hardness ($H_v$) using different formulas. The predicted U-B structures satisfy mechanical stability on the basis of Born stability criteria [63]. As listed in Table. 2, the hardness of UB$_3$-$Cmcm$, UB$_3$-$C2/m$, UB$_4$-$C2/m$ and UB$_6$-$Im$-3 exceed 20 GPa at ambient pressure, in particular UB$_3$-$Cmcm$ (29.59 GPa) and UB$_3$-$C2/m$ (27.39 GPa). The hardness are comparable to NaB$_4$ [64], KB$_7$, SrB$_7$ [24] and CeB$_n$ [26], illustrating the predicted U-B compounds are potential high hardness materials.

TABLE. 2. The bulk modulus $B$ (GPa), shear modulus $G$ (GPa), Young's modulus $E$ (GPa) and the Vickers hardness ($H_v$) estimated by different formulas [52-53] of the predicted U-B compounds.

| | Pressure (GPa) | $B$ (GPa) | $G$ (GPa) | $E$ (GPa) | $\sigma$ | $H_v^{Chen}$ (GPa) | $H_v^{Tian}$ (GPa) | $H_v^{Mazhnik}$ (GPa) |
|---|---|---|---|---|---|---|---|---|
| UB$_3$-$Cmcm$ | 0 | 218.60 | 177.98 | 419.97 | 0.18 | 29.59 | 28.55 | 25.75 |
| UB$_3$-$C2/m$ | 0 | 215.87 | 169.62 | 403.24 | 0.19 | 27.39 | 26.50 | 23.20 |
| UB$_4$-$C2/m$ | 0 | 218.13 | 164.43 | 394.23 | 0.20 | 25.43 | 24.74 | 21.09 |
| UB$_6$-$Im$-3 | 0 | 245.38 | 169.18 | 412.69 | 0.22 | 23.04 | 22.80 | 19.62 |
| UB$_6$-$Cmcm$ | 0 | 235.66 | 150.68 | 372.62 | 0.24 | 19.28 | 19.27 | 17.14 |
| UB$_7$-$Cmcm$ | 60 | 453.27 | 306.82 | 751.00 | 0.22 | 33.11 | 34.03 | 35.22 |
| UB$_8$-$C2/m$ | 20 | 286.20 | 160.94 | 406.61 | 0.26 | 16.93 | 17.45 | 19.41 |
| UB$_{12}$-$C2/m$ | 40 | 328.55 | 231.18 | 561.78 | 0.22 | 29.01 | 29.10 | 27.25 |

### C. Boron clathrates

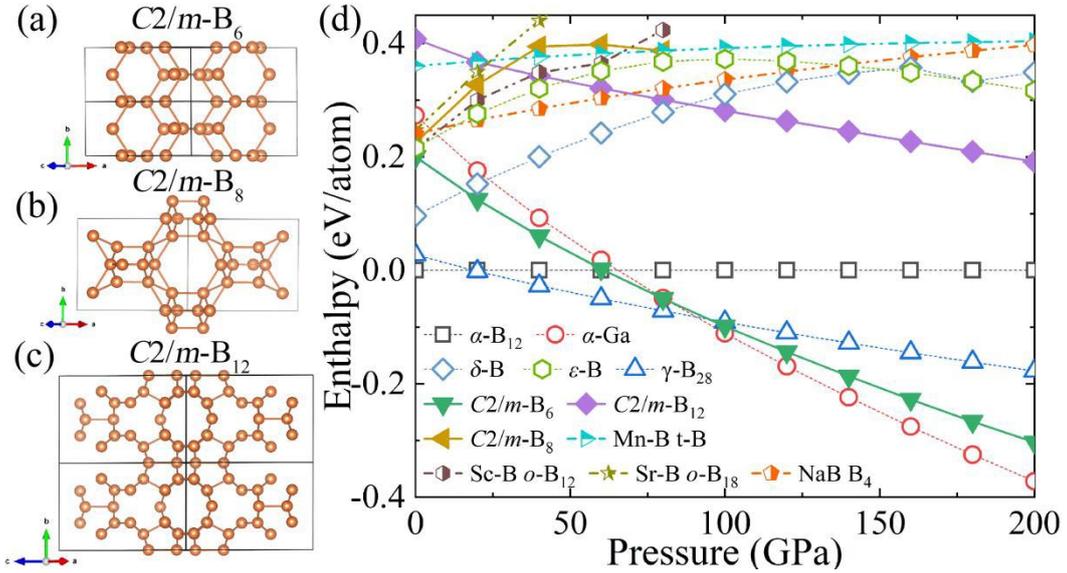

FIG. 4. (a)-(c) The crystal structures of the predicted boron clathrates at ambient pressure. (d) The enthalpy difference of the boron allotropes relative to $\alpha$-B$_{12}$. The t-B, $o$-B$_{12}$, $o$-B$_{18}$ and B$_4$ are predicted structures in Ref. 22, 27-29, $\delta$-B and $\varepsilon$-B are experimentally synthesized meta-stable boron allotropes in Ref. 65.

Since the predicted U-B compounds are made up of the boron networks or cages,

they may act as precursors for unique boron allotropes. By removing the U atoms in the predicted U-B compounds, we optimized the boron structures and calculated their phonon spectra under different pressures. 4 boron clathrate structures are dynamically stable at ambient pressure, including $C2/m$-$B_6$ from $UB_3$-$C2/m$, $I4/mmm$-$B_4$ from $UB_4$-$C2/m$, $C2/m$-$B_8$ from $UB_8$-$C2/m$ and $C2/m$-$B_{12}$ from $UB_{12}$-$C2/m$. Among them, $I4/mmm$-$B_4$ has been reported with a $T_c$ of 19.8 K at ambient pressure [27]. As depicted in Fig. S9 [39], the predicted boron clathrates $C2/m$-$B_6$ and $C2/m$-$B_{12}$ can host their dynamic stability until 200 GPa, while the structure of $C2/m$-$B_8$ collapses after 60 GPa. The crystal structures of $C2/m$-$B_6$, $C2/m$-$B_8$ and $C2/m$-$B_{12}$ are in Fig. 4 (a)-(c), the boron layers in Fig. 3(b) bond with each other after the removal of the U atoms, which lead to the clathrate structures of $C2/m$-$B_6$. The predicted $C2/m$-$B_8$ and $C2/m$-$B_{12}$ retain the boron structure characters of their corresponding compounds [Fig. 3 (g) and (h)]. We calculated the enthalpy differences of $C2/m$-$B_6$, $C2/m$-$B_8$, $C2/m$-$B_{12}$ and the boron allotorpes reported by experiments and theoretical predictions [22, 27-29, 54, 55, 62] relative to the known $α$-$B_{12}$ under high pressures. As shown in Fig. 4 (d), although the enthalpy indicates that $C2/m$-$B_6$ is meta-stable when comparing with the $α$-$B_{12}$ at ambient pressure, $C2/m$-$B_6$ has lower enthalpy value than the synthesized meta-stable boron allotrope of $ε$-B at 0 GPa and $δ$-B at 20 GPa, respectively [65]. With further compression, the enthalpy value of $C2/m$-$B_6$ is lower than the reported $α$-$B_{12}$ after 60 GPa and $γ$-$B_{28}$ around 90 GPa, which indicates the synthesizing potential. Besides, the enthalpy of $C2/m$-$B_8$ is around 0.010 eV/atom above the synthesized $ε$-B at 0 GPa, and the enthalpy of $C2/m$-$B_{12}$ approaches that of $ε$-B with increasing pressure, which is lower than that of $ε$-B after 50 GPa. In the aspect of synthesizing, we propose that one possible route is the combination of precursors and thermal degassing under dynamical vacuum, such as the synthesis of Si-24 from $Na_4Si_{24}$ [66], the other route is analogous to the meta-stable boron allotropes of both $δ$-B and $ε$-B. They are synthesized from $β$-B with the region of pressures 7.5-18 GPa and temperatures of 1373-2373 K [65] using diamond anvil cell (DAC) and laser heating. This suggests that approapriate choice of raw materials and pressure-temperature (P-T) condition are key factors for meta-stable boron allotropes. Considering the relative enthalpy value of the synthesized boron allotropes and other predicted boron structures (t-B, $o$-$B_{12}$, $o$-$B_{18}$ and $B_4$) [22, 27-29] in Fig. 4(c), we assume that $δ$-B with pressure range of 10-20 GPa, $ε$-B around 0 GPa and $ε$-B around 50 GPa could be suitable for $C2/m$-$B_6$, $C2/m$-$B_8$ and $C2/m$-$B_{12}$, respectively.

Then we calculated the electronic structures of the predicted boron clathrates. The

ELF of the predicted boron clathrates are shown in Fig. S10 [39], the boron atoms have strong covalent interactions and the empty region reflect the channels of the clathrate structures in $C2/m$-$B_6$, $C2/m$-$B_8$ and $C2/m$-$B_{12}$. As for the band structures and PDOS of the predicted boron clathrates at ambient pressure plotted in Fig. S11 [39], $C2/m$-$B_6$ is a semi-conductor with a indirect band gap of about 0.88 eV. In $C2/m$-$B_6$, the B-$s$ electrons have competitive contribution with B-$p$ electrons in the valence bands around the Fermi energy, implying the visible hybridization between B-$s$ and B-$p$ in $C2/m$-$B_6$, while $C2/m$-$B_8$ and $C2/m$-$B_{12}$ are typical metals and the B-$p$ electrons play predominant role around the Fermi energy. Moreover, the calculations of COHP in Fig. S12 [39] illustrate that the B-B exhibits bonding characteristics below the Fermi energy in all the three predicted boron clathrates and their ICOHP strengths are comparable with boron allotropes such as $o$-$B_{16}$ [29].

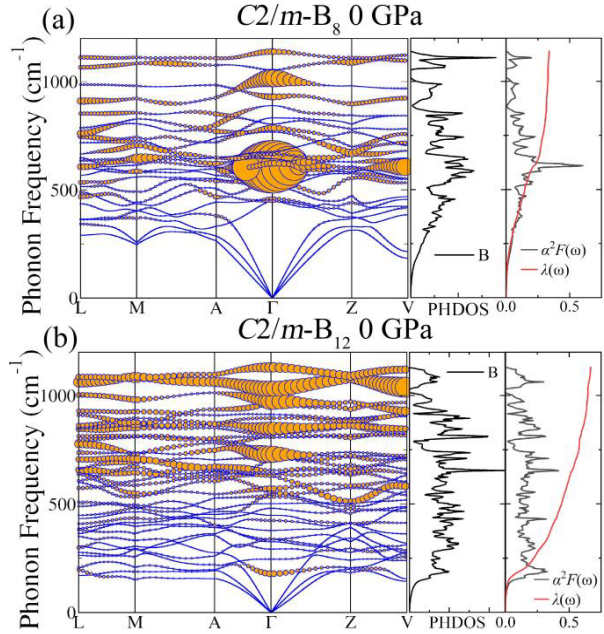

FIG. 5. The calculated phonon curves, PHDOS, Eliashberg spectral function $\alpha^2F(\omega)$, the electron-phonon integral $\lambda(\omega)$ for $C2/m$-$B_8$ and $C2/m$-$B_{12}$ at ambient pressure, respectively. The orange solid dots represents the EPC constant $\lambda$ and the radii are proportional to the strength.

Considering the metallicity of $C2/m$-$B_8$ and $C2/m$-$B_{12}$ at ambient pressure, we carried out the EPC calculations to study their superconducting properties. The calculated EPC constants, projected phonon density of states (PHDOS), Eliashberg spectral function $\alpha^2F(\omega)$ and the electron-phonon integral $\lambda(\omega)$ for $C2/m$-$B_8$ and $C2/m$-$B_{12}$ are shown in Fig. 5, and the radii of the orange dots denote the EPC strength of the vibration modes. In the aspect of vibration modes, frequencies larger than 300 cm$^{-1}$ are more decorated with the solid dots, which correlates with more than 70% of the integral $\lambda(\omega)$ in $C2/m$-$B_8$ [Fig. 5 (a)], and the large distribution of the phonon modes around 500-600 cm$^{-1}$, such as at $\Gamma$ point, result in a sudden enhancement about 29% of the integral $\lambda(\omega)$ [Fig. 5 (a)]. In EPC calculations of $C2/m$-B$_{12}$ [Fig. 5 (b)], we can observe EPC strength distribution around 220 cm$^{-1}$ at $\Gamma$ point, which relates to a peak in $\alpha^2F(\omega)$, and about 29% of the integral $\lambda(\omega)$ is from the vibration modes below 220 cm$^{-1}$, indicating the impact of the relatively low frequencies to the EPC constant. The calculated EPC constant $\lambda$ is 0.34 for $C2/m$-B$_8$ at ambient pressure with a estimated $T_c$ value of 0.96 K ($\mu^*$=0.1), suggesting weak superconductivity. Nevertheless, the EPC constant $\lambda$ is 0.66 for $C2/m$-B$_{12}$ at ambient pressure and the estimated $T_c$ value is 16.12 K ($\mu^*$=0.1). This is comparable to the superconducting boron allotropes like $c$-B$_{24}$ (13.8 K) [67], $I4/mmm$-B$_4$ (19.8 K), $Pm$-B$_{17}$ (15.4 K) [27] and $o$-B$_{16}$ (14.2 K) [29] at ambient pressure.

TABLE. 3. The bulk modulus $B$ (GPa), shear modulus $G$ (GPa), Young's modulus $E$ (GPa) and the Vickers hardness ($H_v$) estimated by different formulas [49-51] of the predicted boron clathrates.

|  | $B$ (GPa) | $G$ (GPa) | $E$ (GPa) | $\sigma$ | $H_v^{Chen}$ (GPa) | $H_v^{Tian}$ (GPa) | $H_v^{Mazhnik}$ (GPa) |
|---|---|---|---|---|---|---|---|
| $C2/m$-B$_6$ | 252.36 | 267.84 | 593.54 | 0.11 | 53.44 | 51.53 | 49.39 |
| $C2/m$-B$_8$ | 188.15 | 118.66 | 294.15 | 0.24 | 16.07 | 16.02 | 13.53 |
| $C2/m$-B$_{12}$ | 151.48 | 58.923 | 156.48 | 0.33 | 4.19 | 5.64 | 8.55 |

Subsequently, we calculated the mechanical properties of the predicted clathrates $C2/m$-B$_6$, $C2/m$-B$_8$ and $C2/m$-B$_{12}$ at ambient pressure, and they satisfy the mechanical stability on the basis of Born stability criteria [63]. As listed in Table. 3, the hardness values of $C2/m$-B$_8$ and $C2/m$-B$_{12}$ are below 20 GPa, while the value of $C2/m$-B$_6$ reaches around 49-53 GPa using different formula [52-54], which reaches the threshold of super-hardness [14, 15] and is comparable to the super-hard borides such as ReB$_2$ (48 GPa), CrB$_4$ (48 GPa), $\alpha$-BeB$_6$ (46 GPa), WB$_5$ (45 GPa) and WB$_4$ (43.3 GPa) [16, 17, 20, 21]. Moreover, we investigated the stress-strain relationships of the super-hard $C2/m$-B$_6$ under various loading conditions. As depicted in Fig. 6 (a), the tensile strain calculations are along the [001], [100], [010], [101], [110], [001] and [111] directions. The ideal tensile strength is about 28 GPa, owning to the (110) plane is the easy cleavage plane, and the tensile strength along the other directions are larger than 30 GPa. Following the tensile strain calculations [Fig. 6(a)], we studied the shear stress by loading the strain on the easy cleavage plane (110) in Fig. 6 (b). The shear directions along [1$\bar{1}$0] and [1$\bar{1}\bar{1}$] have isotropic respond to the shear strain within the $\varepsilon$=0.1. Different from the other two shear directions, there is a sharp decrease under the deformation of $\varepsilon=0.15$, suggesting the weakest shear stress or the ideal shear strength of 23 GPa along the [001] direction of (110) plane.

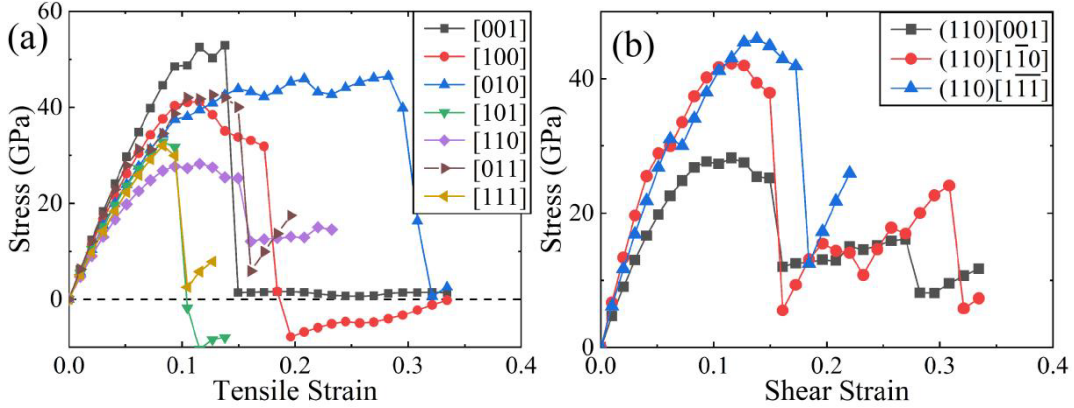

FIG. 6. Stress-strain relations of $C2/m$-$B_6$ under (a) tensile and (b) shear deformations.

To further validate the thermodynamic stability of $C2/m$-$B_6$ and $C2/m$-$B_{12}$, we performed the AIMD (*NVT* ensemble) simulations under ambient pressure at 300 K and 500 K. The simulation results along with the intermediate structures in Fig. S13 (a) and (b) [39] illustrate that $C2/m$-$B_6$ and $C2/m$-$B_{12}$ could withstand their stability under ambient condition without obvious structural destruction, suggesting that $C2/m$-$B_6$ and $C2/m$-$B_{12}$ have the potential to be synthesized under appropriate experimental condition. In addition, we further carried out the AIMD (*NVT* ensemble) simulations for the precursors of UB$_3$-$C2/m$ and UB$_{12}$-$C2/m$ at 500 K under their loweset dynamical stable pressure. As shown in Fig. S13(c) [39], the precursor structures could preseve their stability at 500 K as well. Moreover, the compounds and boron structures were rechecked by AIMD (*NPT* ensemble) at 300 K in Fig. S14 [39], which further suggests the dynamical stability at finite temperature. As for the precursors, the UB$_{12}$-$Fm$-$3m$ is one of the experimentally reported structure and the predicted UB$_{12}$-$C2/m$ has lower enthalpy after 167 GPa. Their enthalpy difference is less than 0.1 eV/atom after 120 GPa [Fig. S1] [39]. Hence we propose that UB$_{12}$-$C2/m$ has potential for synthesizing under high pressure and high temperature (HPHT) combing DAC and laser heating around 120 GPa, or we can expect the structure transformation under higher pressure and then release the pressure. In UB$_3$-$C2/m$, we propose to take the experimental $Imm2$-Na$_2$B$_{29}$ as reference [68]. The $Imm2$-Na$_2$B$_{29}$ has positive formation energy of 3.03 meV/atom comparing with Na plus B, which is far from the convex hull [68]. It could be synthesized using sodium and boron, which are filled into a pre-boronated tantalum tube and exposed to reaction conditions (1050 °C, 3 h, plus 1150 °C, 3 h). Then an excess of sodium is removed by distillation at $10^{-2}$ mbar and 350 °C [69]. Furthermore, Ref. 70 points out that uranium borides could be synthesized combining UO$_2$, boron carbide (B$_4$C), graphite and the partial pressures of CO. Thus, we assume that UB$_3$-$C2/m$ is likely to be synthesized when considering the above mentioned experimental clues.

Overall, our calculations provide theoretical evidence to support the existence of the predicted super-hard $C2/m$-B$_6$ and superconducting $C2/m$-B$_{12}$ under certain experimental conditions. The removing of U atoms in the precursors may be one of the access for achieving the predicted boron clathrates, we can also expect the emerging of $C2/m$-B$_6$ and $C2/m$-B$_{12}$ when the experiments meet appropriate requirements during the synthesizing of boron allotopes.

## IV. Conclusion

In summary, using the first-principles calculations and crystal structure prediction methods in MAGUS, we systematically investigated the phase diagram of U-B compounds within 200 GPa. We have proposed 8 unique compounds that are possible to be synthesized and 4 exotic stoichiometries. These predicted compounds have either net or caged structure units which are potential precursors of the boron allotropes. After removing the uranium atoms, we studied the stability, electronic structures, superconductivity and mechanical properties of the corresponding boron structures. We predicted three boron clathrates under ambient pressure, among which $C2/m$-B$_6$ is a super-hard material with the estimated Vickers hardness value about 49-53 GPa, while $C2/m$-B$_{12}$ is superconducting with the $T_c$ value of 16.12 K. $C2/m$-B$_6$ and $C2/m$-B$_{12}$ have potential for synthesizing at ambient condition and they are potential materials for future applications. Our exploration enriches the phase diagram of U-B compounds and boron, which is helpful for designing unique and safer material for the nuclear industry and provides insights for studies on the unique boron based materials.

## Acknowledgement

This work was supported by the National Natural Science Foundation of China (Grant Nos. 12474018, 52272265, 12204138 12125404 and 123B2049), the National Key R&D Program of China (Grant Nos. 2023YFA1607400 and 2022YFA1403201), the Basic Research Program of Jiangsu (Grant BK20233001, BK20241253), the Jiangsu Funding Program for Excellent Postdoctoral Talent (Grants 2024ZB002 and 2024ZB075), the Postdoctoral Fellowship Program of CPSF (Grant GZC20240695), the AI & AI for Science program of Nanjing University, and the Fundamental Research Funds for the Central Universities. The calculations were supported by HPC